\documentclass[11pt,english,a4paper]{article}
\usepackage{babel}
\usepackage{amsmath,amsfonts}
\usepackage{graphicx}
\usepackage[all]{xy}

\def\Tr{\mathrm{Tr}}
\def\IR{\mathbb{R}}
\def\half{\frac{1}{2}}

\title{Global properties of an exact string theory solution in two and
  four dimensions}  
\author{Harald G. Svendsen
\\
Albert-Einstein-Institut}
    
\begin{document}
\begin{flushright}
AEI-2005-174
\end{flushright}
\bigskip
\begin{center} 
  {\Large \bf Global properties of an exact string}
  \bigskip
  {\Large \bf theory solution in two and four dimensions} 
\end{center}

\bigskip\bigskip

\centerline{\bf Harald G. Svendsen} 

\bigskip\bigskip
\centerline{\it Max-Planck-Institut f{\"u}r Gravitationsphysik} 
\centerline{\it (Albert-Einstein-Institut)} 
\centerline{\it Am M{\"u}hlenberg 1}
\centerline{\it D-14476 Potsdam-Golm, Germany}
\centerline{\small \tt harald.svendsen@aei.mpg.de}

\bigskip\bigskip

\begin{abstract}
This paper discusses global properties of exact (in $\alpha'$) string
theory solutions: A deformed black hole solution in two dimensions and
a Taub-NUT type solution in four dimensions.  These models are exact
by virtue of having CFT descriptions in terms of heterotic coset models. 
The analysis includes analytic continuations of the metric,
motion of test particles, and the T-duality which acts as a map
between different regions of the extended solutions, rendering the
physical spacetimes non-singular.
\end{abstract}


\section{Introduction}

An interesting extension of ordinary coset models are the heterotic
coset models \cite{Johnson:1994kv}, where
%
fermions are included in a left/right non-symmetric fashion with
supersymmetry only in the right-moving sector. These fermions contribute
to the total anomaly, which has to cancel for consistent theories. 
In the usual coset models, anomaly cancellation essentially restricts
the allowed gaugings to the vector ($g\to hgh^{-1}$) and axial 
($g\to hgh$) choices \cite{Bars:1991pt}. In the heterotic construction,
however, the left-moving fermions have arbitrary couplings to the
gauge fields, and by tuning these couplings, we are allowed a wider
range of  gaugings. 

Starting with a heterotic
$SL(2,\mathbb{R})\times SU(2) / [U(1)\times U(1)]$
model, the associated spacetime geometry in the low-energy
approximation has been computed some time ago \cite{Johnson:1994jw},
and shown to correspond to the throat region of a stringy Taub-NUT
solution \cite{Johnson:1994ek}.  
%
The \emph{exact} (in $\alpha'$) geometry was recently worked out in
ref.~\cite{Johnson:2004zq}, by noting that 
the entire worldsheet action can be written, after bosonisation of the
fermions, as a sum of gauged Wess-Zumino-Novikov-Witten (WZNW)
models. 
For such models it is relatively easy to write down the quantum
effective action, \cite{Tseytlin:1992ri,Bars:1993zf}, 
in which the fields should be treated as classical fields. 
The gauge fields can then be integrated out in a direct way equivalent
to solving their equations of motion \cite{Witten:1991yr}, and give a
result valid to \emph{all} orders in the relevant parameter. 
Before reading off the background fields, it is necessary to take into
account that the bosonised fermions really are fermions. In principle
we should re-fermionise them, but since we are only interested in the
background metric and dilaton, it is enough to rewrite the action in a
form that prepares it for re-fermionisation
\cite{Johnson:1994ek}. When the action is put in 
this form, we can readily read off the fields.
%
%
%

%
The resulting solution in string frame is \cite{Johnson:2004zq}:
\begin{equation}
\begin{split}
  \label{eq:stringyTaubNUT}
  ds^2 & =  (k-2)\biggl[ \frac{dx^2}{x^2-1} 
    - \frac{x^2-1}{D(x)}(dt-\lambda\cos\theta d\phi)^2 
    + d\theta^2 + \sin^2\!\theta d\phi^2 \biggr],
    \\
  e^{2\Phi} & =  D(x)^{-\frac{1}{2}},
\end{split}
\end{equation}
where
\begin{equation}
    D(x)= (x+\delta)^2-\frac{4}{k+2}(x^2-1),
\end{equation}
and $\delta\geq 1$, $k>2$, $-\infty<x<\infty$, $0<\theta<\pi$, 
$0<\phi<2\pi$, and importantly, the coordinate $t$ is periodic with period
$4\pi\lambda$. 
In the coset model construction, the periodicity of $t$ arises as a
result of the gauging, and is necessary also to avoid a conical
singularity.
A periodic $t$ means that there are closed timelike curves in
the regions $x>1$ and $x<-1$. 
The Einstein frame metric is given by $ds^2_E= e^{-2\Phi}ds^2$ (in 4
dimensions).

The main motivation for performing the computations outlined above,
was that this model provides a 
good laboratory to investigate the fate of closed timelike curves and
cosmological singularities in string theory. It is exact in $\alpha'
\sim \frac{1}{k}$ and shows that the essential features of the low
energy limit ($k\to\infty$) survive to all orders, indicating that
$\alpha'$ corrections are not sufficient to rule out closed timelike
curves. However, there might be other corrections to the solution,
e.g.~string coupling ($g_s$) corrections.
In this paper we will investigate global properties of this
solution, and demonstrate by use of T-duality that the
apparently singular regions disappear in the full solution. This
result agrees with previous investigations of the bosonic
$SL(2,\mathbb{R})/U(1)$ black hole 
\cite{Witten:1991yr,Dijkgraaf:1991ba,Perry:1993ry}.


The 4D stringy Taub-NUT solution \eqref{eq:stringyTaubNUT} has
topology $\mathbb{R}\times S^3$ 
and can be viewed as a fibre bundle over 
$S^2$ with fibre $\mathbb{R}\times S$, just as the Taub-NUT space in
General Relativity \cite{HawkingEllis}.
The fibre can be regarded as the $(x,t$) plane, and its metric is
obtained from \eqref{eq:stringyTaubNUT} by dropping terms in $d\theta$
and $d\phi$, giving
\begin{equation}
  \label{eq:fibrecoverspace}
  ds^2 = (k-2) \biggl[ \frac{dx^2}{x^2-1} - \frac{x^2-1}{D(x)}dt^2
  \biggr]. 
\end{equation}
The covering space (where $t$ is non-compact) of this 2D geometry is
described by a \emph{heterotic} $SL(2,\mathbb{R})/U(1)$ model.
This has been shown in the low-energy ($k\to\infty$) limit
\cite{Johnson:1994jw}, and in the next section we will verify this
statement also for general $k$.
For this reason, the global properties of the stringy Taub-NUT
geometry are the same as those of this 2D solution,
which we will therefore focus on in the following.


\section{Deformed 2D black hole}
\label{sec:2dsolution}


In this section we shall study the $SL(2,\mathbb{R})/U(1)$ heterotic
coset model 
\cite{Johnson:1994jw} and show that its exact metric
is identical to eq.~\eqref{eq:fibrecoverspace} (with non-compact
$t$). It can be viewed as a deformation of the bosonic 2D black hole
first studied in ref.~\cite{Witten:1991yr}, the exact geometry of
which was worked out in ref.~\cite{Dijkgraaf:1991ba}.
The computation summarised in this section is virtually identical to
the one done for the 4D stringy Taub-NUT space in
ref.~\cite{Johnson:2004zq}, and more details and references are given
there.

The action for the bosonic sector is a gauged WZNW action,
\begin{equation}
  \label{eq:gWZNWaction}
  S = k[I(g) + I(g,A)],
\end{equation}
where $g\in SL(2,\mathbb{R})$. The constant $k$ is in CFT language the
level constant, and should be identified with the string tension,
$k\sim \frac{1}{\alpha'}$. 
The $I(g)$ is an ungauged WZNW action, given as
\begin{equation}
\begin{split}
  \label{eq:ungaugedWZNWaction}
  I(g)=&-\frac{1}{4\pi}\int d^2\!z
  \Tr( g^{-1}\!\partial g g^{-1}\!\bar{\partial}g)
  -i\Gamma(g),
  \\
  \Gamma(g)=&\frac{1}{12\pi}\int \Tr(g^{-1}\!dg)^3,
\end{split}
\end{equation}
where $\Gamma(g)$ is the Wess-Zumino term.
The coupling to the $U(1)$ gauge field is governed by the action
\cite{Witten:1991mm}
\begin{equation}
\begin{split}
  \label{eq:gWZNWgaugecoupling}
  I(g,A)= \frac{1}{4\pi}\int d^2\!z \Tr\Bigl(&
  2 \bar{A}^R g^{-1} \partial g
  -2 A^L \bar{\partial} g g^{-1} 
  +2 A^L g \bar{A}^R g^{-1} 
  \\ &
  +(A^L\bar{A}^L + A^R\bar{A}^R)
  \Bigr) ,
\end{split}
\end{equation}
where $A\equiv A_z$ and $\bar{A} \equiv A_{\bar{z}}$ are the
gauge field components, $T^L$ are the left 
acting generators, $T^R$ are the right acting generators, 
and we have used the notation
$A^L = A_a T^Ldz^a = A^L_{a}dz^a$ and
$A^R = A_a T^R dz^a = A^R_{a}dz^a$, for $a=\{z,\bar{z}\}$.

Let us parametrise the group elements according to
\begin{equation}
\begin{split}
  g_b &=e^{t_L\sigma_3/2} e^{r\sigma_1/2} e^{t_R\sigma_3/2} 
  \\
  &=\frac{1}{\sqrt{2}}\begin{pmatrix}
    e^{t_+/2}(x+1)^{1/2}      & e^{t_-/2}(x-1)^{1/2}
    \cr e^{-t_-/2}(x-1)^{1/2} & e^{-t_+/2}(x+1)^{1/2}
  \end{pmatrix} \in SL(2,\IR)\ ,
  \label{eq:parameters1}
\end{split}
\end{equation}
where $t_\pm=t_L\pm t_R$, and $-\infty\leq t_R, t_L \leq\infty$, and
$x=\cosh r$.
Although $x$ is introduced in this way, we can allow it to take any
real value while remaining in $SL(2,\IR)$.
Note that if $-1<x<1$, then $g_b$ is of the form
$\bigl(\begin{smallmatrix} a & ib \\ ic & d \end{smallmatrix}\bigr)$
where $a,b,c,d\in \mathbb{R}$, 
and by the isomorphism given as $t_-\to t_-+i\pi$
this matrix is isomorphic to 
$\bigl(\begin{smallmatrix} a & -b \\ c & d \end{smallmatrix}\bigr)$,
which is a real  $SL(2,\mathbb{R})$ matrix.
Similarly, if $x<-1$, then $g_b$ is of the form
$\bigl(\begin{smallmatrix} ia & ib \\ ic & id \end{smallmatrix}\bigr)$
which by the isomorphism $t_\pm\to t_\pm+i\pi$ is
isomorphic to
$\bigl(\begin{smallmatrix} -a & -b \\ c & d \end{smallmatrix}\bigr)$,
which again is a real  $SL(2,\mathbb{R})$ matrix.

For the model to have (0,1) worldsheet supersymmetry, we need  two
right-moving fermions, minimally coupled to the gauge field with unit
charge.
For the left-moving sector we are free to add two left-moving fermions
which are coupled to the gauge field with an arbitrary coupling $Q$.
After bosonisation, these fermionic degrees of freedom are represented
by one bosonic field $\Phi$, whose action is a gauged WZNW model 
of the form of eq.~\eqref{eq:gWZNWaction}, but now based on
the group $SO(2)$ at level $k=1$ \cite{Witten:1983ar}.
Let us parametrise this sector according to
\begin{equation}
  g_f = e^{ \Phi i\sigma_2/\sqrt{2}}
    = \left( \begin{array}{cc}
          \cos\frac{\Phi}{\sqrt{2}} & \sin\frac{\Phi}{\sqrt{2}} \\
          -\sin\frac{\Phi}{\sqrt{2}} & \cos\frac{\Phi}{\sqrt{2}}\\
        \end{array} \right) \in SO(2).
    \label{eq:parametersf}
\end{equation}
Our heterotic model is thus equivalent to a purely bosonic coset
model based on $SL(2,\mathbb{R})_k\times SO(2)_1 / U(1)$.
The level constant $k$ is related to the central charge $c$ by
\begin{equation}
  c=\frac{3k}{k-2}-1+1,
\end{equation}
where the first term is from $SL(2,\mathbb{R})$, the $-1$ is from
gauging $U(1)$, and $+1$ is the fermionic contribution.
To cancel the conformal anomaly we could add an
appropriate internal CFT, but this will not be relevant for our
investigations. 


The gauging of the subgroup $U(1)$ is implemented as
\begin{equation}
  \label{eq:gauging}
  g\to h_L g h_R, 
  \qquad h_L=e^{\epsilon T^L},
  \quad h_R = e^{\epsilon T^R},
\end{equation}
where the generators $T^{L,R}$ are
\begin{equation}
\begin{aligned}
  \label{eq:generators}
  T^L_b &= \frac{\sigma_3}{2}, 
  & 
  T^R_b &= \delta\frac{\sigma_3}{2},
  \\
  T^L_f &= -Q\frac{i\sigma_2}{\sqrt{2}},
  & 
  T^R_f &= -\delta\frac{i\sigma_2}{\sqrt{2}}.
\end{aligned}
\end{equation}
Since we are interested in geometries with Lorentzian signature, we
want to gauge a noncompact subgroup of the $SL(2,\mathbb{R})$
part.
Our notation is admittedly sloppy, in that we think of the gauge
subgroup as a ``noncompact $U(1)$'' $\cong SO(1,1)$ when acting
on the $SL(2,\mathbb{R})$ part, but as a proper $U(1)\cong SO(2)$ when
acting on the $SO(2)$ part. We use this notation for simplicity, and
the reader should not attach any further significance to 
this.

%
%
For generic choice of parameters, this gauging gives an action
that is not gauge invariant even at the classical level. For this
gauge anomaly to cancel, the anomalous contributions from the bosonic 
and fermionic sector have to cancel, giving us the equation
\begin{equation}
  k(\delta^2-1)+2(\delta^2-Q^2)=0.
\end{equation}
Once this equation is satisfied, it follows that also the quantum
effective action is anomaly free.
This anomaly cancellation condition makes it clear that we can chose
arbitrary gauging parameter $\delta$ by adjusting the coupling
constant $Q$.
%
%
We choose the gauge fixing condition $t_L=0$, and will in the
following write $t_R=t$. 


With this setup, the exact metric and dilaton can be computed as
outlined in the introduction.
The metric is the same as in eq.~\eqref{eq:fibrecoverspace} (with
non-compact $t$),
%
and the dilaton is as in the 4D solution \eqref{eq:stringyTaubNUT}.
The coordinates $\{x,t\}$ can both take any real value.
For $\delta=1$ this is the same as the exact solution of the purely
bosonic $SL(2,\mathbb{R})/U(1)$ model \cite{Dijkgraaf:1991ba} (with
the replacement $\frac{4}{k+2}\to \frac{2}{k}$ in the function
$D(x)$). 

The metric \eqref{eq:fibrecoverspace} has Killing horizons at $x=\pm
1$ and curvature singularities where $D(x)=0$, which happens for
$x=x^c_\pm$, given by 
\begin{equation}
  x^c_\pm =-\frac{(k+2)\delta \pm 2\sqrt{\delta^2(k+2)-(k-2)}}{k-2}.
\end{equation}
For $\delta>1, k>2$ we always have $x^c_\pm<-1$.
%
Note that for $\delta\neq \pm 1$, the $SL(2,\mathbb{R})$ symmetry
transformations $g\to h_Lgh_R$ and $g\to h_Lgh_R^{-1}$ have no fixed
points, and the discussion in
refs.~\cite{Dijkgraaf:1991ba,Ginsparg:1992af} relating metric
singularities with fixed points therefore does not apply.
%
%
The (string frame) Ricci curvature scalar is
\begin{equation}
  R = -\frac{1}{(k-2)D^2}\Bigl[ (x^2-1)2D D'' -3(x^2-1)(D')^2 + 6x DD',
  \Bigr] 
\end{equation}
where prime denotes differentiation with respect to $x$.
It is finite at $x=\pm 1$, but diverges at $x^c_\pm$.

The geometry can be divided into different regions with different
properties. 
There is an asymptotically flat region for $x>1$ (region I).
A Killing horizon at $x=1$ connects it to an interior region $-1<x<1$
(region II).  
There is another Killing horizon at $x=-1$ which connects to a new
region $x^c_+<x<-1$ (region III). 
The region between the curvature singularities, $x^c_-<x<x^c_+$
(region IV), has Euclidean signature and an imaginary dilaton.
On the other side of the singularity, $x<x^c_-$ (region V), there is
another asymptotically flat region. 
We will comment on  analytic continuations of the metric in the next
section.

Taking the low-energy approximation ($k\to\infty$) has the effect that
$x^c_-\to x^c_+$ such that the Euclidean region (region IV)
disappears, leaving one curvature singularity at $x^c=-\delta<-1$. 
On the other hand, sending $\delta\to 1$ has the effect that $x^c_+\to
-1$ such that  region III disappears. The curvature singularity
associated with $x^c_+$ vanishes, leaving a mere boundary between a
Lorentzian region (region II) and a Euclidean region (region IV).
If we take both limits, both regions III and IV disappears, leaving a
curvature singularity at $x=-1$.
%


By continuing to Euclidean time, $t\to i\theta$, and writing $x=\cosh
r$ it is easy to compute the Hawking temperature associated with the
horizon at $x=1$ ($r=0$).  In the neighbourhood of $r=0$ the Euclidean
line element becomes
\begin{equation}
  ds^2 \simeq (k-2)\Bigl[ dr^2 + r^2
  \bigl(\frac{d\theta}{1+\delta}\bigr)^2 \Bigr].
\end{equation}
To avoid a conical singularity at $r=0$ it is necessary for Euclidean
time $\theta$ to have periodicity $2\pi(1+\delta)$.
The Hawking temperature defined as the inverse of the proper length of
the Euclidean time at infinity is 
$T_H = \bigl( 2\pi(1+\delta)\sqrt{k+2}\bigr)^{-1}$.


\section{Global structure}

In this section we will study analytic continuations of the metric and 
show that geodesics can be extended past the horizons at $x=\pm 1$. We
will also investigate the motion of test particles, which demonstrates
that the singularities are shielded by potential barriers.
The global structure of the closely related bosonic
$SL(2,\mathbb{R})/U(1)$ (essentially this is the special case where
$\delta=1$) solution has been discussed in
refs.~\cite{Dijkgraaf:1991ba,Perry:1993ry}.

\begin{figure}
\centering
\includegraphics{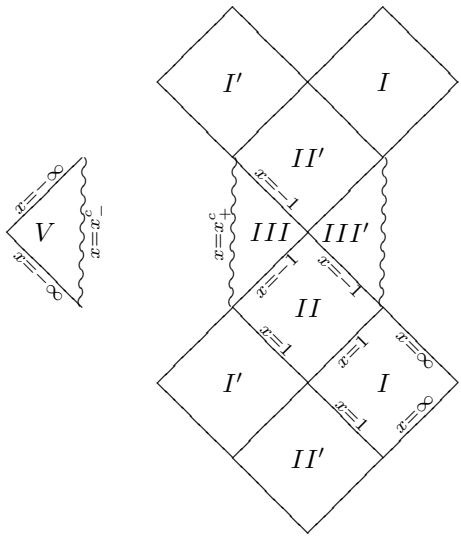}
  \caption{\label{fig:penroseextended}Penrose diagram for extended
  solution. The diagram continues indefinitely in the vertical
  direction. The singularities in regions $III$, $III'$ and $V$ are all
  repulsive, and massive particles approaching these singularities
  bounce back.}
\end{figure}

%
The metric as written down in eq.~\eqref{eq:fibrecoverspace} is singular at
$x^c_\pm$ and at the horizons $x=\pm 1$. The bad behaviour at the
curvature singularities $x^c_\pm$  can of course not be resolved by a
change of coordinates, but we will now 
demonstrate that there are analytic continuations across the
horizons at $x=\pm 1$ by rewriting the metric in terms of null
coordinates. 
Consider a null curve with affine parameter $\tau$. The tangent
vector $u^a=\frac{d x^a}{d\tau}\equiv \dot{x}^a$ satisfies
$0=u^2=g_{ab}\dot{x}^a\dot{x}^b$, which gives
\begin{equation}
  \label{eq:nullequation}
  dt = \pm \frac{D(x)^{\half}}{x^2-1} dx.
\end{equation}
Define null coordinates $(u,v)$ which satisfy
\begin{equation}
  du = dt - \frac{D(x)^\half}{x^2-1}dx,
  \qquad
  dv = dt + \frac{D(x)^\half}{x^2-1}dx.
\end{equation}
Using the first of these relations, we can write the metric
\eqref{eq:fibrecoverspace} as
\begin{equation}
  \label{eq:2dmetric_u}
  ds^2 = -(k-2)D(x)^{-\half} \biggl[\frac{x^2-1}{D(x)^\half}du + 2dx\biggr]du.
\end{equation}
In these coordinates the line element is well defined for $x=\pm 1$,
showing explicitly that these are mere coordinate singularities. It
also demonstrates that the metric can be straightforwardly continued
from region I to II and to III.
Our use of the coordinate $x$ already in the parametrisation of
$SL(2,\mathbb{R})$ in eq.~\eqref{eq:parameters1} of course anticipated
this result. 
The Penrose diagram for the maximally extended spacetime is shown in
figure \ref{fig:penroseextended}.
The metric can also be written in double null coordinates,
\begin{equation}
  ds^2 = -\frac{x^2-1}{D(x)} du dv,
\end{equation}
where $x$ should now be thought of as a function of $u$ and $v$.

Null geodesics are null curves and therefore satisfy
eq.~\eqref{eq:nullequation},
where the $\pm$ represent two different families of null geodesics.
The equations diverge as $x\to \pm 1$, but this is again due to the
bad choice of coordinates. 
In the coordinates $(u,x)$ of eq.~\eqref{eq:2dmetric_u} the equations
for the two families of null geodesics become 
\begin{equation}
  u=\text{const.} 
  \quad \text{or} \quad
  \frac{du}{dx}=-\frac{D(x)^\half}{x^2-1},
\end{equation}
which shows that the first family can be continued across $x=\pm 1$. 
The second equation, representing the second family of null geodesics,
still diverges as $x\to\pm 1$, but this divergence can be avoided in
the same way by working with the coordinates $(v,x)$. This is just
like the situation in the 2D Misner or 4D Taub-NUT solutions of
General Relativity \cite{HawkingEllis}.

Now, let us study the motion of test particles in this geometry.
To slightly simplify the expressions we absorb the overall factor
($k-2$) of the metric by a rescaling of the coordinates.
Consider particles which couple to the string frame metric (but not to
the dilaton), with the Lagrangian
\begin{equation}
  L = \frac{1}{2}g_{ab}\dot{x}^a\dot{x}^b
  = \frac{1}{2}\frac{1}{x^2-1}\dot{x}^2 
  -\frac{1}{2}\frac{x^2-1}{D(x)}\dot{t}^2.
\end{equation}
The dot represents differentiation with respect to the affine parameter
$\tau$.
The existence of the Killing vector $\xi=\frac{\partial}{\partial t}$
leads to the conserved quantity
\begin{equation}
  p_t = -g_{ab}\xi^a\dot{x}^b = \frac{x^2-1}{D(x)}\dot{t}.
\end{equation}
Using this together with the identity
$g_{ab}\dot{x}^a\dot{x}^b=\kappa$, where $\kappa=0,\pm 1$,
we can compute the trajectory of the particles.

\begin{figure}
  \centering
  \includegraphics{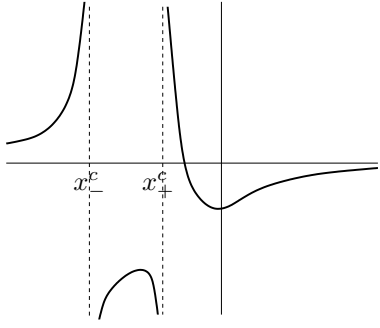}
  \caption{\label{fig:potential}
    Potential $V(x)$. Asymptotically, it goes to zero as 
    $x\to\pm\infty$.}
\end{figure}

For \emph{massive} particles (timelike trajectory) we have
$\kappa=-1$, which gives  
\begin{equation}
  \label{eq:geodesic}
  \dot{x}^2 = D(x)\bigl[ p_t^2-\frac{x^2-1}{D(x)} \bigr]
  = 2 D(x)\bigl[ E - V(x) \bigr],
\end{equation}
where $E=\frac{1}{2}(p_t^2-\beta_0)$ is the particle's energy
(constant), and $V(x)=\frac{1}{2}(\frac{x^2-1}{D(x)}-\beta_0)$ is the
potential. The constant $\beta_0=\frac{k+2}{k-2}$ is chosen such that
$V(\infty)=0$.  Note that as long as we are outside the Euclidean
region, $D(x)>0$, with $D(x)\to 0$ as $x\to x^c_\pm$. The potential is
plotted in figure \ref{fig:potential}.
Differentiation of equation \eqref{eq:geodesic} gives 
(if $\dot{x}\neq 0$) 
\begin{equation}
  \label{eq:acceleration}
  \ddot{x} = \frac{\partial}{\partial x}\Bigl[ D(x)[E-V(x)] \Bigr]
  =ax+b,
\end{equation}
where
\begin{equation}
  a=2E\frac{k-2}{k+2}, \quad
  b=\delta(2E+\frac{k+2}{k-2}).
\end{equation}
This can easily be solved to give
\begin{equation}
  \label{eq:testparticlesolution}
  x(\tau) = -\frac{b}{a}+\Bigl(\frac{b}{a}+x_0\Bigr)\cosh(\sqrt{a}\tau)
  +\frac{v_0}{\sqrt{a}}\sinh(\sqrt{a}\tau),
\end{equation}
with energy $E$ related to the inital position $x_0=x(0)$ and
velocity $v_0=\dot{x}(0)$ of the particle by
\begin{equation}
  E=\frac{v_0^2}{2D(x_0)}+V(x_0).
\end{equation}
The solution \eqref{eq:testparticlesolution} is for $E>0$. If $E<0$,
the hyperbolic functions have to be replaced by their trigonometric
cousins, and $a$ with the absolute value $|a|$.

Assume that we start with a particle coming in from the right 
($x>0$, $\dot{x}<0$). 
Initially, the potential decreases, so that $E-V$ increases and
$\dot{x}^2$ remains positive. The particle can pass through the
horizons at $x=1$ and $x=-1$, but will then meet the potential
barrier. For some value $x^c_+<x<-1$, we have $E-V\to 0$, and
since $D$ is finite, 
$\dot{x}^2\to 0$, while $\ddot{x}>0$. In other words, the particle is
reflected by the \emph{repulsive} singularity at $x^c_+$.
A similar repulsion of massive particles happens also in other
solutions, e.g. in Reissner-Nordstr{\"o}m spacetime in the
$GM^2<p^2+q^2$ case.

For \emph{massless} particles (null trajectory) we have $\kappa=0$,
giving eqs.~\eqref{eq:geodesic} and \eqref{eq:acceleration} with
$V(x)=0$ and $\beta_0=0$. 
The solution is the same as for the massive case
\eqref{eq:testparticlesolution}, but now with  
$b=2 \delta E$.
%
The differences however, have a significant effect on the
physics. Since a massless particle sees no potential barrier,
it can come in from positive $x$ and travel all the
way into the singularity at $x^c_+$, with finite value of the affine
parameter. This means that there are incomplete null geodesics due
to the curvature singularity.

In the special case $\delta=1$ (studied in 
refs.~\cite{Dijkgraaf:1991ba,Perry:1993ry}) we have a
qualitatively different situation although the solution
\eqref{eq:testparticlesolution} is still valid.
In this case the point $x^c_+=-1$ is a boundary between Lorentzian and
Euclidean signature, but is \emph{not} a curvature singularity. 
The potential $V(x)$
is smooth and increasing for $x>x^c_-$, which would suggest that
particles can pass through the $x=-1$ horizon, enter the Euclidean
region and eventually fall into the singularity at $x^c_-$. 
But we do not expect it to be possible for a particle to enter a
region with different signature, so the question is what really
happens at $x=-1$.   
If we view this spacetime as the limiting case $\delta\to 1$ we find
that the potential barrier associated with the singularity $x=x^c_+$
approaches $x=-1$ and becomes infinitely steep. And following the
discussion above (for $\delta>1$), the point where a particle is reflected
approaches $x=-1$ (independently of the particle's energy).
In some form we expect this argument to hold also ``after the limit''
when $\delta=1$.
And indeed it does.
If $\delta=1$ there is no potential barrier anymore to stop an
incoming particle, but as $x\to -1$ we have
$D(x)\to 0$. Noting that $E-V>0$ is finite, we get from
eq.~\eqref{eq:geodesic} that $\dot{x}^2\to 0$ and since $\ddot{x}$ is
positive we can conlude that the particle is reflected at $x=-1$. 
In this sense the repulsive singularity at $x^c_+$ disappears, but
leaves behind a ``boundary of reflection''.
The reflective nature of the boundary $x=-1$ has been known for a long
time \cite{Dijkgraaf:1991ba,Perry:1993ry}, but the point we want to
make here is that this can be understood as a limiting case of the
$\delta>1$ models, where the reflection is easily understood as a
potential barrier effect due to a repulsive singularity.

If in addition to $\delta=1$ we let $k\to\infty$ (solution studied in
ref.~\cite{Witten:1991yr}), the 
Euclidean region disappears, and there is a true curvature singularity
at $x=-1$ (since also $x^c_-\to -1$) shielded by a horizon and thus
representing a conventional black hole.
As $x\to -1^+$ we have 
$\half\dot{x}^2=D(x)[E-V(x)]\to -D(x)V(x)\to x^2-1 \to 0$
and 
$\ddot{x}=2E(x+1)+1\to 1$.
Since the point $x=-1$ now is a curvature singularity we conclude
that massive particles in this case hit the singularity, which is
of course normal black hole behaviour.



\section{T-Duality}
\label{sec:duality}

It was noted already in the original paper on the bosonic
$SL(2,\mathbb{R})/U(1)$ black hole \cite{Witten:1991yr} that there is a
duality in the solution, corresponding to choosing either vector and
axial gauging, which are the two anomaly free gaugings in that model.
In ref.~\cite{Dijkgraaf:1991ba} this duality was discussed further in
the context of the $\alpha'$ exact solution. and certain
generalisations of it has been discussed in
refs.~\cite{Bars:1991pt,Bars:1992ti}. 

One may wonder whether this duality is special to axial and vector
gauging, and it is not immediately clear how it extends to the
heterotic case ($\delta\neq 1$). Now we don't have the same notion of
axial ($g\to hgh$) and vector ($g\to hgh^{-1}$) gauging, but gauging
given by $g\to h_Lgh_R$, where $h_L$ and $h_R$ are 
related in a profoundly non-symmetric way,
cf. eq.~\eqref{eq:generators}. 
In this section we shall investigate this question explicitly and
demonstrate that the duality indeed is there. 
We will show that it amounts to changing the sign of right-moving
currents, and is given by the transformation
$h_R\leftrightarrow h_R^{-1}$, resembling the usual axial/vector
duality. 
Our discussion follows to a 
large extend ref.~\cite{Kiritsis:1991zt}, but is generalised to the
case of asymmetric left/right gaugings. 

Let us start with a general derivation of a duality for scalar fields
$\phi$ coupled to some $\phi$ independent current $J$.
Consider the action
\begin{equation}
	\label{eq:motheraction}
	S[A,B^a] = \int d^2\!z \Bigl[
	B\bar{B} + i(B\bar{\partial}A-\bar{B}\partial A) 
	-2(B\bar{J}-\bar{B}J) - 2J\bar{J}
	\Bigr],
\end{equation}
where $J\equiv J_z$ and $\bar{J}\equiv J_{\bar{z}}$ are independent of
$A$ and $B^a$.
Integrating out the field $A$ in the partition function
$Z=\int \mathcal{D}A\mathcal{D}B\exp{(-S[A,B^a])}$ produces a delta
function $\delta(\bar{\partial}B-\partial \bar{B})$. 
This delta function means that we should 
introduce a scalar variable $\phi$ and define $B=\partial\phi$,
$\bar{B}=\bar{\partial}\phi$.  
This change of variables gives a trivial Jacobian and makes the delta
function integrate to unity. The partition function then becomes
$Z=\int\mathcal{D}\phi\exp{(-S_A[\phi])}$ where
\begin{equation}
	\label{eq:daughteractionA}
	S_A[\phi] = \int d^2\!z \Bigl[
	\partial\phi\bar{\partial}\phi
	+2(J\bar{\partial}\phi-\bar{J}\partial\phi) -2J\bar{J}
	\Bigr].
\end{equation}
If on the other hand we integrate out the field $B^a$ from
eq.~\eqref{eq:motheraction} the partition function becomes
$Z=\int\mathcal{D}\phi\exp{(-S_B[\phi])}$ where 
\begin{equation}
	\label{eq:daughteractionB}
	S_B[\phi] = \int d^2\!z \Bigl[
	\partial\phi\bar{\partial}\phi
	+2(J\bar{\partial}\phi+\bar{J}\partial\phi) +2J\bar{J}
	\Bigr],
\end{equation}
and the scalar field $\phi$ is now defined as $\phi=iA$.
The two actions \eqref{eq:daughteractionA} and
\eqref{eq:daughteractionB} are dual to each other, and by simple
inspection we see that the transformation between them is done by
sending $\bar{J}\to -\bar{J}$.

We want to show that our model exhibits a duality of this form. To do
this we will rewrite the gauged WZNW action \eqref{eq:gWZNWaction} in
a form comparable to \eqref{eq:daughteractionA}. We can then read off
$J$ and $\bar{J}$, and see what effect the duality transformation has
by studying what it means to send $\bar{J}\to-\bar{J}$. 
Let us factorise the field $g\in SL(2,\mathbb{R})$ according to
\begin{equation}
	g= hQ, \qquad Q = e^{i\phi T_0} \in U(1),
\end{equation}
where $iT_0$ is a generator of $U(1)$ normalised as $\Tr(T_0
T_0)=1$. We consider cases where $T_0$, $T_R$ and $T_L$ commute. 
Using the Polyakov-Wiegmann formula
\cite{Polyakov:1983tt,Polyakov:1984et}, 
\begin{equation}
	I(ab) = I(a) + I(b) 
	-\frac{1}{4\pi}\int d^2\!z \bigl[
	2\Tr(a^{-1}\partial a \, \bar{\partial} b b^{-1}) \bigr],
\end{equation}
where $I(g)$ is defined in eq.~\eqref{eq:ungaugedWZNWaction},
we can write the gauged WZNW action \eqref{eq:gWZNWaction} as
\begin{equation}
\begin{split}
	S = kI(h) + \frac{k}{4\pi}\int d^2\!z \Bigl[ &
		\partial\phi\bar{\partial}\phi
		+2\bigl[ (-i)(U_0+AM_{L0})\bar{\partial}\phi 
		- (-i)\bar{A}P_R\partial\phi \bigr]
		\\ &+2\bar{A}U_R - 2A\bar{U}_L + 2A\bar{A}(M_{LR}
		+\half X)
	\Bigr],
\end{split}
\end{equation}
where we have defined
\begin{equation}
\begin{aligned}
  \label{eq:shorthandnotation}
	U_0 &= \Tr(T_0 h^{-1}\partial h ),
	&P_{R} &= \Tr(T_0 T_R),
	\\
	U_R &= \Tr(T_R h^{-1} \partial h),
	& M_{L0} &= \Tr(T_L h T_0 h^{-1}),
	\\
	\bar{U}_L &= \Tr(T_L \bar{\partial} h h^{-1}),
	& M_{LR} &= \Tr(T_L h T_R h^{-1}),
	\\
	X &= \Tr(T_LT_L+T_RT_R).
\end{aligned}
\end{equation}
If we define
\begin{equation}
  \label{eq:quasicurrents}
	J = -i(U_0+AM_{L0}), \qquad
	\bar{J} = -i\bar{A} P_R,
\end{equation}
and note that 
$U_0P_R=U_R$ and $M_{L0}P_R=M_{LR}$,
then the action takes the form
\begin{equation}
\begin{split}
	S = kI(h) + \frac{k}{4\pi}\int d^2\!z \Bigl[ &
		\partial\phi\bar{\partial}\phi
		+2(J\bar{\partial}\phi - \bar{J}\partial\phi) -2J\bar{J}
		-2\bar{A}\bar{U}_L + A\bar{A} X)
	\Bigr].
\end{split}
\end{equation}
This action is of the form of eq.~\eqref{eq:daughteractionA}, and we
therefore know immediately that the dual action is found by taking
$\bar{J}\to -\bar{J}$.
From the definitions \eqref{eq:shorthandnotation} and
\eqref{eq:quasicurrents}, we see that this is equivalent to
$T_R\to -T_R$, which again means $h_R\to h_R^{-1}$ as promised at the
outset. 

The duality action on the dilaton follows immediately from these
considerations. Recall that the dilaton is given by the determinant
appearing when we integrate out the gauge fields, while
the duality we have demonstrated above relates actions before any
integration of gauge fields has been performed. The dual 
dilaton is therefore given as the determinant arising from the dual
action. 

The currents associated with the global  $G_L\times G_R$ symmetry of
the ungauged WZNW model are
\begin{equation}
  j = \Tr(\partial g g^{-1} T^L),
  \qquad
  \bar{j} = \Tr(g^{-1}\bar{\partial} g T^R).
\end{equation}
So the duality $T_R\to -T_R$ of our model corresponds to changing the
sign of the right-moving current, $\bar{j}\to -\bar{j}$, just as we
would expect from a T-duality.
For the case where $T_R=\pm T_L$, this is of course nothing but the
well-studied vector/axial duality of coset models.

Having demonstrated the existence of such a duality in the CFT of
non-symmetrically gauged models, the interesting question is how it
works in terms of the background fields.
As already remarked, this is simply done by
replacing $T_R$ with $-T_R$ in all the
calculations, and the resulting metric and dilaton are found to be the
same as before, providing we take $x\to -x$.
This means that the duality transformation maps between the regions
$x<0$ and $x>0$ of the extended solution, in agreement with what has
been observed before
\cite{Witten:1991yr,Dijkgraaf:1991ba,Bars:1991pt,Bars:1992ti,Perry:1993ry}.

\begin{figure}
  \centering
  \includegraphics{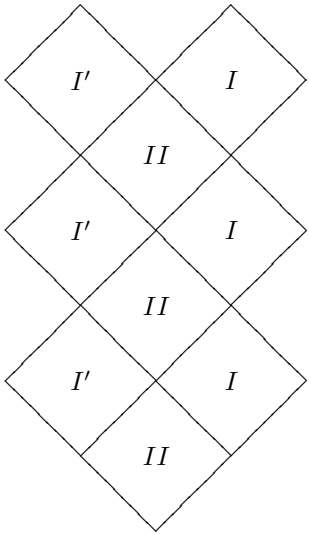}
  \caption{\label{fig:penroseduality}Penrose diagram for extended
  non-singular solution. The diagram continues indefinitely in the vertical
  direction.}
\end{figure}

This duality means that the singularities we have found in the $x<0$
region are indeed artifacts of the description, as the region is dual to the
non-singular $x>0$ region. Remember that the dilaton blows
up near the singularities, and hence the string coupling $g_s$ becomes
strong there. It is natural to assume that the singularities
arise as a result of ignoring strong string coupling effects. In the
$x>0$ region we do not have this problem of $g_s$ becoming strong, so
the solution can be trusted. We therefore conclude that the geometry is
everywhere non-singular. The Penrose diagram for this extended
physical geometry is shown in figure \ref{fig:penroseduality}.

\section{Stringy Taub-NUT}

The interesting part of the 4D stringy Taub-NUT solution
\eqref{eq:stringyTaubNUT} is the fibre represented by the coordinates
$(t,x)$, the metric of which is identical to the 2D solution studied
thus far \eqref{eq:fibrecoverspace}.
So the entire analysis of the previous sections can be carried over
directly to the stringy Taub-NUT solution.
The base space is topologically an $S^2$ with little interesting
physics attached to it, and the only relevant difference from the 2D
solution is that the $t$ coordinate now has periodicity
$4\pi\lambda$. This gives rise to 
closed timelike curves in the NUT regions $I$, $I'$, $III$, $III'$ and
$V$ of the Penrose diagram in figure \ref{fig:penroseextended}. The
existence of closed timelike curves in the exact solution is a very
interesting property, indicating that $\alpha'$ corrections are not
sufficient to resolve the problems associated with such curves.
This issue is one of the main motivations for studying this model, but
not something we discuss any further in the present paper.

When studying the motion of test particles in this solution, one could
argue that it should be done in the Einstein frame rather than in
string frame, because the graviton and dilaton don't mix there.
This can be done, and the conclusion that particles bounce off the
singularity at $x^c_+$ due to a potential barrier still holds. 
However, unlike the string frame metric, the Einstein frame metric is
not asymptotically flat.

A sign change in the right-moving gauge
group generators, $T_R \leftrightarrow -T_R$, 
manifests itself in the spacetime metric and dilaton of the 4D model
as the transformation $(x,\phi) \leftrightarrow (-x,-\phi)$, which is
therefore the effect of a T-duality transformation in the present
case.

A rotating generalisation of the stringy Taub-NUT spacetime was
constructed in ref.~\cite{Johnson:1995ga}, and the $\alpha'$ exact
geometry was computed in ref.~\cite{Svendsen:2004fi} following the
procedure outlined in section \ref{sec:2dsolution}. This model has an
extra constant $\tau$ which parametrises the rotation, such that
$\tau=0$ gives back the non-rotating solution
\eqref{eq:stringyTaubNUT}. 
Non-zero $\tau$ breaks the global $SU(2)$ rotational symmetry of the
stringy Taub-NUT solution down to an axial symmetry, represented by the
Killing vector $\partial_{\phi}$.
For small values of $\tau$ (under-rotating case) the solution is a
smooth deformation of the $\tau=0$ case, while for $\tau$ larger than
some critical value (over-rotating case) the loci $x^c_\pm(\theta)$ of
the curvature singularities are deformed so much that their topology
changes. For small $\tau$ they appear one outside the other, centred
around the origin $x=0$ and appearing in the
negative $x$ region, while for $\tau$ large they form ``bubbles''
outside the origin, one in
the negative $x$ region, and one in the positive $x$ region. See
ref.~\cite{Svendsen:2004fi} for details.

In the under-rotating case we can refer to the $x\leftrightarrow -x$
duality and conclude that since the $x>0$ region is non-singular, the
entire geometry is non-singular.
In the over-rotating case, however, the duality is not enough to
demonstrate that the geometry is really non-singular, since neither of
the regions are themselves free of curvature singularities.
Nevertheless, we should keep in mind that the singularities represent
loci where the string coupling becomes infinite, and the solution as
written down cannot be trusted where this happens. 
We expect, in analogy with the $\tau=0$ case, that the
rotating $\tau\neq 0$ solution is also non-singular, although we
cannot prove this directly  with the help of T-duality.

\section{Summary}

In this paper we have presented the full $\alpha'$ corrections of a
2-dimensional solution of string theory which
is identical to the fibre part of the exact four-dimensional stringy
Taub-NUT spacetime.

We have studied global properties of these solutions,
and have discussed analytic continuations of the solution and
presented the Penrose diagram for the extended spacetime.
An investigation of test particles in the geometry showed that massive
particles approaching the singularities are repelled by a potential
barrier, while massless particles hit the singularity with finite
value of the affine parameter.
The perfect reflection boundary $x=-1$, which is a feature of 
the exact bosonic $SL(2,\mathbb{R})/U(1)$ black hole, can in this
context be understood heuristically as a result of taking the limit
$\delta\to 1$ where the potential barrier becomes infinitely steep and
localised near 
$x=-1$. 

The axial/vector duality which exists in coset models was generalised
to the case of heterotic coset models, where the left and right gauge
actions are asymmetric in a non-trivial way. This T-duality was then
used to resolve the curvature singularities, rendering the 4D stringy
Taub-NUT and 2D deformed black hole spacetimes non-singular.

\subsection*{Acknowledgements}
Thanks to J\"urg K\"appeli for comments on the
manuscript. 
This work is supported by the Alexander von Humboldt Foundation (Germany).

\bibliographystyle{JHEP}
\bibliography{refs}

\end{document}